# Hedonic Adaptation in the Age of AI: A Perspective on Diminishing Satisfaction Returns in Technology Adoption

Venkat Ram Reddy Ganuthula[1]     Krishna Kumar Balaraman[1]

Nimish Vohra[1]

Indian Institute of Technology Jodhpur

**Abstract**

The fast-paced progress of artificial intelligence (AI) through scaling laws connecting rising computational power with improving performance has created tremendous technological breakthroughs. These breakthroughs do not translate to corresponding user satisfaction improvements, resulting in a general mismatch. This research suggests that hedonic adaptation—the psychological process by which people revert to a baseline state of happiness after drastic change—provides a suitable model for understanding this phenomenon. We argue that user satisfaction with AI follows a logarithmic path, thus creating a long-term "satisfaction gap" as people rapidly get used to new capabilities as expectations. This process occurs through discrete stages: initial excitement, declining returns, stabilization, and sporadic resurgence, depending on adaptation rate and capability introduction. These processes have far-reaching implications for AI research, user experience design, marketing, and ethics, suggesting a paradigm shift from sole technical scaling to methods that sustain perceived value in the midst of human adaptation. This perspective reframes AI development, necessitating practices that align technological progress with people's subjective experience.

**Keywords:** Artificial Intelligence; Hedonic Adaptation; User Satisfaction; Scaling Laws; Technology Adoption

## 1. Introduction

We are now at the point in the development of artificial intelligence (AI) according to scaling laws that has provided us with an age of unmatched technological advancement. This milestone means that improved performance can be consistently achieved through higher investments in computational power, data, and model parameters (Kaplan et al., 2020). The underlying growth laws found have enabled phenomenal advancements across a spectrum of domains—extending to language processing (Hoffmann et al., 2022), computer vision (Zhai et al., 2022), and multimodal systems (Radford et al., 2021)—which in turn has enabled the development of advanced models like GPT-3 (Brown et al., 2020) and its derivatives. This trend, backed by power-law scaling relations between capability and utilization of resources (Henighan et al., 2020), is the basis of our collective and individual endeavors at breaking new shores in artificial intelligence, as witnessed through mounting investments to fuel better benchmarks. But with these advances, there unfolds a paradox: technical capabilities are advancing at a breakneck speed, yet that does not necessarily translate to increasing user satisfaction, and thus poses a challenge to our present understanding of the social impact of AI (Schmidt & Cohen, 2017; Brynjolfsson et al., 2018). In this work, we submit that hedonic adaptation—the psychological process by which individuals revert to a baseline level of satisfaction despite veritable changes (Brickman & Campbell, 1971)—is a necessary framework to explain this disconnect, and thus we explore the interplay between AI potentiality and human experience.

Our interest in scaling laws is because they are empirically valid, which indicates that performance can be quantified with respect to model size, data size, and computational power (Kaplan et al., 2020). This trend has enabled the development of high-end capabilities—tasks that were hitherto impossible (Wei et al., 2022)—as witnessed through the shift from rudimentary algorithms to advanced multimodal generative models (Radford et al., 2021). Both industry and academia have chased the model, investing progressively larger architectural scales in the hope that increases in capability will equate to increases in value (Brown et al., 2020). But a more subtle interpretation of user feedback offers a different explanation. Dramatically enhanced AI accuracy or utility—such as those witnessed in conversational agents (Leviathan & Matias, 2018)—tend to yield little more than incremental increases in user satisfaction or retention, a pattern that has garnered increasing attention from industry observers (Schmidt & Cohen, 2017; Brynjolfsson et al., 2018). This disconnect prompts the question of why a stable correlation between technical excellence and a gain in subjective valuation does not exist, compelling us to look beyond the quantitative measures that make up our discipline.

We argue that hedonic adaptation, a long-established construct in psychological theory, offers a reasonable explanation for the phenomenon. First described by Brickman and Campbell (1971), hedonic adaptation—the phenomenon more commonly known as the "hedonic treadmill"—demonstrates how individuals quickly adapt to large positive and negative events, thus sustaining stable levels of happiness despite external fluctuations. The phenomenon has been widely documented across a variety of settings: for example, increases in income ultimately lose emotional importance (Easterlin, 1995), the psychological impact of disability fades gradually over time (Oswald & Powdthavee, 2008), and initial excitement for new consumer products ultimately becomes the norm (Frederick & Loewenstein, 1999). Lyubomirsky (2011) briefly describes the phenomenon, suggesting that "people's emotional systems adjust to their current circumstances and recalibrate their neutral point" (p. 187), suggesting that subjective gains are short-lived as reference points shift. Applied to technology, the trend is observed: people rapidly adapt new capabilities into their expectations, resulting in declining satisfaction even as performance gains are achieved (Norton et al., 2012). With artificial intelligence, this trend seems to be amplified by the speed of advancement—capabilities that are initially deemed to be remarkable, like the conversational ability of Google Duplex, are promptly accepted as the standard, with customers being remorseful of restrictions instead of expressing continued awe (Leviathan & Matias, 2018, p. 14).

This process of adaptation compels us to re-examine the benchmarks by which we judge the success of artificial intelligence, because psychological studies reveal that faster change accelerates shifts in reference points (Lyubomirsky, 2011). Adoption research on technology also corroborates this assertion, as it posits that satisfaction in users typically erodes after an initial phase as novelty gives way, even where systems are found to be reliable (Bhattacherjee & Premkumar, 2004). Kim and Malhotra (2005) also observe that experience is a future judgment anchor, a process seen in AI settings where use of tools like ChatGPT fades as novelty declines (Polyportis, 2024). The "hype cycle" outlined by Gartner (Fenn & Raskino, 2008) is an apt analogy, mapping out technologies from the periods of hyperbolic expectation to subsequent disillusion, though it lacks the mechanistic explanation afforded by adaptation theory. We contend that user satisfaction is logarithmic in relation to capability—a concept aligned with Weber-Fechner psychophysical laws, where perception grows logarithmically with the intensity of stimulus (Dehaene, 2003). This implies that the initial transition from primitive to intermediate capability can generate as much satisfaction as a subsequent transition from high to extraordinary performance, though the latter has more technical merit, thereby opening up a long-lasting "satisfaction gap" defying the presumed linear nature of subjective returns (Sheldon et al., 2013).

Our analysis propounds four central arguments, each grounded in this theoretical synthesis. First, we posit that hedonic adaptation as a mediator of AI capability translation into user satisfaction requires re-examining priorities in development. Psychological research confirms that adaptation is an ineluctable process (Frederick & Loewenstein, 1999), and with AI—where progress is accelerating, as evidenced by the GPT-3 to GPT-4 jump (Bodonhelyi et al., 2024)—this means that users readjust expectations with every technological breakthrough, finally modulating satisfaction (Lyubomirsky, 2011). Second, we posit that the dynamics of satisfaction occur in discrete phases: initial quick wins as capabilities surmount lower baselines, returns diminishing as reference points rise, stabilization as progress becomes normal, and sporadic jumps induced by breakthrough innovations (Fenn & Raskino, 2008). This is in alignment with known cycles of adoption, where early adoption peaks before dipping (Polyportis, 2024). Third, we propose that drivers like the speed of capability introductions and heterogeneity of users affect these phases, offering possibilities for intervention. Klapwijk and Van Lange (2009) hold that satisfaction is reference-point dependent, so that the manner of breakthrough delivery influences the rate of adaptation, while heterogeneous user pools—early and late adopters—respond in different ways (Rogers, 2003). Finally, we believe that filling this gap redesigns strategies regarding AI development, design, market competition, and policy and requires transition from naive scaling to strategies that maintain subjective worth (Brynjolfsson et al., 2018).

These hypotheses force us to reconcile psychological understanding with technical aspirations, acknowledging that AI's scaling laws (Kaplan et al., 2020) paint a portrait of exponential growth of capability, but human adaptation brings this down to a logarithmic subjective experience (Dehaene, 2003). Our use of technical measures—e.g., cross-entropy loss or accuracy—ignores this mediation, for research shows human judgment plateaus before measures (Clark et al., 2021; Li et al., 2022). Industry trends illustrate this tension: despite heavy investment, user retention trails (Schmidt & Cohen, 2017), implying that experience must be given priority over capability in isolation. We structure this view as follows: Section 2 surveys key literature, Section 3 outlines our conceptual framework, Section 4 offers theoretical justification, Section 5 presents practical implications, Section 6 addresses limitations and future directions, and Section 7 offers closing remarks. By combining psychology's understanding of adaptation (Lyubomirsky, 2011), adoption's emphasis on user dynamics (Rogers, 2003), and AI's technical path (Kaplan et al., 2020), we aim to redefine AI's development—not as a quest for technical supremacy, but as a challenge to navigate human adaptation, so that innovations speak as deeply as they compute.

**2. Theoretical Foundations**

The accelerated arrival of artificial intelligence (AI) over recent years, fueled by scaling laws guaranteeing repeated improvement in performance with added computation, data, and model parameters (Kaplan et al., 2020), compels close inspection of the human experience of such technological progress. Dramatic developments in language models (Hoffmann et al., 2022), computer vision (Zhai et al., 2022), and multimodal systems (Radford et al., 2021) signal technical mastery in AI, but residual disconnect persists: user satisfaction doesn't grow in proportion to capability, as reports in industry attest to minimal subjective advancement while actual progress has crystallized (Schmidt & Cohen, 2017; Brynjolfsson et al., 2018). We claim that hedonic adaptation—the psychological process whereby humans recover a settled level of contentment despite catastrophic change (Brickman & Campbell, 1971)—supplies the fundamental framework to account for disconnect. Drawing upon ideas from psychological adaptation theory, technology adoption theories, and research in AI scaling, this section presents the theory underlying a model where satisfaction plots a logarithmic curve, thereby opening a "satisfaction gap" challenging prevalent paradigms in AI advancement. Such underpinnings, founded in considerable literature, elucidate interactions between technological advancement and human experience, thus allowing a reassessment of AI's role in society.

Hedonic adaptation, originally codified by Brickman and Campbell (1971), proposes that people rapidly adapt to large-scale life events and revert to a relatively stable baseline of happiness irrespective of change size. This has been rigorously supported across a range of environments: gains in income lose their affective kick over time (Easterlin, 1995), distress of impairment declines with adaptation (Oswald & Powdthavee, 2008), and thrill over new possessions wears off to be mundane (Frederick & Loewenstein, 1999). Lyubomirsky (2011) describes the process, writing, "The positive effects of a change are usually transient because people's emotional systems adjust to their current circumstances and recalibrate their neutral point" (p. 187). This adaptation, commonly referred to as the "satisfaction treadmill" (Kahneman & Thaler, 2006), contends that objective gains have difficulty attaining subjective well-being as reference points move upwards. In technology settings, this force is no less prevalent. Sheldon et al. (2013) showed that happiness from new technology follows a pattern of diminishing returns, proposing that users rapidly internalize innovation into expectations. Klapwijk and Van Lange (2009) also contend that satisfaction is highly reference-dependent, with prior experience anchoring subsequent judgments, a process which seems particularly strong in AI, where capabilities change at an unprecedented rate. Leviathan and Matias (2018) captured this in their work on conversational AI, writing that "users appear to normalize extraordinarily quickly to capabilities that would have seemed magical just months prior, subsequently focusing criticism on remaining limitations rather than achieved capabilities" (p. 14), detecting adaptation's applicability to areas of rapid change.

Technology adoption theory offers a competing explanation, illustrating how subjective dynamics unfold beyond the uptake initiation stage. Rogers' (2003) diffusion of innovations theory characterizes dimensions like perceived usefulness, ease of use, and social influence as drivers of adoption, a theoretical framework Davis' (1989) Technology Acceptance Model (TAM) builds on to include perceptions of usability. Subsequent revisions added emotional and experiential elements (Venkatesh et al., 2012; Yang & Forney, 2013); however, these models are focused mainly on adoption choices rather than usage. Bhattacherjee and Premkumar (2004) offer valuable findings by showing user attitudes and perceptions shift following adoption, with satisfaction falling from initial levels even as system performance lives up to expectations—a finding that suggests expectations are dynamically recalibrated over time. Kim and Malhotra (2005) concur, finding judgments are strongly reference dependent, forming dynamic reference points that change with experience. This finding is consistent with Gartner's "hype cycle" (Fenn & Raskino, 2008), descriptively capturing the path that technologies follow in passing through hype, disillusion, and ultimately, productivity; although, it is short on psychological detail of adaptation. These tendencies forecast a phenomenon by which novelty around AI

tools—demonstrated through ChatGPT's initial uptake (Polyportis, 2024)—ultimately wears off to allow normalization, a process consistent with the hedonic adaptation's modulating influence on subjective gains (Frederick & Loewenstein, 1999).

The technical underpinnings of AI advances are based on scaling laws, which offer a quantitative model for capability increase but reveal a subjective chasm when perceived through human eyes. Kaplan et al. (2020) showed language model performance to scale as a power-law function of model size, dataset size, and computational budget, increasing as $L(N) \propto N^{-\alpha}$ with an exponent $\alpha$ usually between 0.05-0.10—a relationship cross-validated on topics from reinforcement learning (Hoffmann et al., 2022) to vision transformers (Zhai et al., 2022) to multimodal systems (Henighan et al., 2020). Those improvements are realized as emergent capabilities on hard tasks (Wei et al., 2022), fueling investment in ever-larger models such as GPT-3 (Brown et al., 2020). But the leap from technical measurement to user-conceived quality isn't so obvious. Diaz and Madaio (2023) criticize scaling laws for being grounded in assessment metrics out of sync with multifarious user opinion, contending that "performance gains promised by increasing dataset size and model scale frequently fail to convert to user-perceived quality in diverse populations" (p. 2). Li et al. (2022) showed human ratings of language model output scale sub-linearly with model size, while Clark et al. (2021) showed satisfaction with AI assistants plateaus well short of technical milestones, indicating psychological mediators to that translation. Both are consistent with hedonic adaptation's hypothesis that rapid gains in capability—however dazzling—may be quickly normalized, for users adapt to each new metric (Lyubomirsky, 2011).

This synthesis of theory points to a key omission: although psychology, adoption, and scaling literatures each touch on facets of AI interaction with users, few integrate them to examine adaptation's role in shaping capability-satisfaction dynamics—a gap of interest considering AI's velocity and social stakes. We hypothesize hedonic adaptation produces an inherent disconnection, wherein capabilities increase through power laws ($C(r) \propto r^\alpha$) but satisfaction follows a logarithmic path ($S(C) \propto k \cdot \log(C) + b$), with k and b depending on context. This model, based on Weber-Fechner logarithmic perception principles (Dehaene, 2003), predicts doubling capability yields approximately equal subjective increases, irrespective of starting point—a pattern where a jump from rudimentary to moderate capability might be equal to the thrill of a subsequent leap from high to exceptional, even though the latter represents a larger technical leap (Sheldon et al., 2013). This "satisfaction gap" contradicts conventional strategies of relentless scaling, as observed with industry advancements (Brynjolfsson et al., 2018), and resonates with adoption studies where interest declines in the aftermath of hype (Polyportis, 2024).

Other theories offer partial analogs but not such an integrative view. Expectation-Confirmation Theory (ECT) (Oliver, 1980; Bhattacherjee, 2001) links satisfaction to differences in performance expectation and discrete events vs. continuous adaptation. Novelty-Seeking Theory (Hirschman, 1980; Coombs & Avrunin, 1977) links declining returns to habituation but not exponential growth of AI. The Unified Theory of Acceptance and Use of Technology (UTAUT) (Venkatesh et al., 2003) explains initial adoption via performance and effort expectancies, but post-adoption evolution. The Peak-End Rule (Kahneman et al., 1993) dwells on peak and end, reversing adaptation's focus on ongoing reference shifts (Table 1). Hedonic adaptation, however, captures AI's power-law growth and logarithmic satisfaction as dynamic, long-term interaction, offering a robust foundation for reassessing strategy development (Kaplan et al., 2020). This synthesis—accepting psychology's view on human adjustment (Lyubomirsky, 2011), adoption's focus on dynamics of users (Rogers, 2003), and scaling's technical expertise—underlies the claim that AI's future is not just a matter of capability but of recognizing human capacity to evolve with it.

**Table 1: Comparison of Theoretical Frameworks for Understanding Technology Satisfaction**

| Theoretical Framework | Primary Mechanism | Temporal Focus | Mathematical Formalization | Key Predictions |
|---|---|---|---|---|
| Hedonic Adaptation Model (Current paper) | Reference point shifts based on experienced capabilities | Continuous post-adoption trajectory | $S(C) = k \cdot \log(C) + b$; Reference point updating: $R_i(t+1) = R_i(t) + \gamma_i \cdot (C(t) - R_i(t))$ | Logarithmic relationship between capabilities and satisfaction; diminishing returns from repeated exposure; adaptation rate determines satisfaction decline slope |
| Expectation-Confirmation Theory (Oliver, 1980; Bhattacherjee, 2001) | Disconfirmation between expectations and | Pre-adoption to initial post-adoption | Satisfaction = f(Performance - Expectations) | Binary satisfaction outcomes (confirmed/disconfirmed expectations); emphasis on initial adoption |

| | perceived performance | | | decision rather than ongoing usage |
|---|---|---|---|---|
| Novelty-Seeking Theory (Hirschman, 1980; Coombs & Avrunin, 1977) | Inherent psychological tendency toward habituation | General personality trait affecting adoption | Not formally specified in technology contexts | Early adoption behavior; technology switching patterns; preference for novel features over improvements to existing ones |
| Unified Theory of Acceptance and Use of Technology (UTAUT) (Venkatesh et al., 2003) | Behavioral intentions based on performance expectancy, effort expectancy, social influence, and facilitating conditions | Adoption decision | Usage Intention = $\beta_1$(Performance Expectancy) + $\beta_2$(Effort Expectancy) + $\beta_3$(Social Influence) + $\beta_4$(Facilitating Conditions) | Primary focus on adoption likelihood rather than satisfaction trajectory; stable preferences rather than dynamically updated ones |

| Peak-End Rule (Kahneman et al., 1993) | Evaluations based on peak intensity and final moments rather than average experience | Retrospective evaluation of completed experiences | Overall Evaluation ≈ (Peak Experience + End Experience)/2 | Disproportionate influence of capability peaks and most recent experiences on overall satisfaction; limited account of continuous reference point updating |
|---|---|---|---|---|
| Gartner Hype Cycle (Fenn & Raskino, 2008) | Market-level enthusiasm followed by disillusionment and eventual productivity | Technology lifecycle at market level | Descriptive model without formal mathematical specification | Predictable phases of market enthusiasm and disappointment; focuses on collective rather than individual satisfaction dynamics |

## 3. Conceptual Model of Hedonic Adaptation in AI Adoption

The rapid development of artificial intelligence (AI) presents a one-time chance to explore the interaction between human psychological processes and technological advancements, in this case, through the framework of hedonic adaptation. This section presents a theoretical model of the processes of adaptation by users to AI capabilities over time, based on proven technology acceptance theory and psychology concepts to operationalize the relationship between objective measures of improving performance and subjective levels of satisfaction. Hedonic adaptation, first proposed by Brickman and Campbell (1971), is the process by which individuals return to baseline levels of satisfaction in response to radical change, a process that has been extensively replicated in a range of domains, including changes in income (Easterlin, 1995), physical impairment (Oswald & Powdthavee, 2008), and consumer spending habits (Frederick & Loewenstein, 1999). In AI, this means that users will quickly assimilate new capabilities into expectations, resulting in a decrease in subsequent satisfaction as standards of technology develop (Norton et al., 2012). This theoretical model suggests that satisfaction is subject to fluctuation in a cyclical direction through a dynamic process of interaction between the evolution of AI capabilities and the changing reference points of users, under the influence of a wide range of factors including rates of adaptation, patterns of capability delivery, and heterogeneity of the user population, and hence presents a conceptual basis for the explanation and potentially modulation of this process.

At the core of this theory is the perception of artificial intelligence (AI) capability as a fluid concept, responsive to resource allocations in terms of scaling laws presented in the literature. Kaplan et al. (2020) depicted that AI performance improves in a power-law relationship correlated with computational resources, data quantity, and model parameters; this pattern is observed for most language models (Hoffmann et al., 2022), vision systems (Zhai et al., 2022), and multimodal models (Radford et al., 2021). Such a pattern of progress suggests that capabilities may improve either incrementally, as in the case of sequential model improvement (Brown et al., 2020), or in jumps, as in major drops revealing new capabilities (Wei et al., 2022).

Psychological theory hypothesizes that the pace and nature of such advances affect adaptation of users: rapid improvement may accelerate changes in reference points, as users move emotional standards toward aligning with prevalent states (Lyubomirsky, 2011). For instance, Leviathan and Matias (2018) found that Google Duplex users quickly adapted to receive its conversational capabilities, prioritizing small deficits over chronic awe (p. 14). This suggests the incremental development of AI paradoxically hastens adaptation, reducing the perceived importance of each successive innovation.

User satisfaction, in this model, arises from the interaction between these dynamic capabilities and individual reference points, which dynamically recalculate to represent experienced performance. Frederick and Loewenstein (1999) contend that adaptation arises as individuals integrate new stimuli into baseline expectations, a phenomenon seen in technology applications where initial euphoria gives way as features become the norm (Sheldon et al., 2013). In AI, this could take the form of users marveling at an initial model's capabilities—e.g., GPT-3's text generation (Brown et al., 2020)—only to normalize it as later models like GPT-4 arrive (Bodonhelyi et al., 2024). The speed of this adjustment differs, with psychological studies demonstrating that rapid adaptation hastens the return to baseline satisfaction (Wilson & Gilbert, 2008). Klapwijk and Van Lange (2009) also suggest that satisfaction is reference-dependent, with past experience anchoring future judgments, a phenomenon repeated in longitudinal technology studies where satisfaction falls after adoption despite continued performance (Bhattacherjee & Premkumar, 2004). This model views satisfaction as a function of the difference between capability and reference point, where the emotional response to gains and losses is asymmetric, as prospect theory suggests (Kahneman & Tversky, 1979). Losses below expectation—e.g., an AI not meeting a newly set benchmark—may produce more pungent dissatisfaction than similar gains above it produce delight, suggesting that early bounds create excitement, but later improvements create diminishing returns.

This asymmetry is in line with general psychological principles, namely the Weber-Fechner law, whereby subjective experience grows logarithmically with regard to intensity of stimulus (Dehaene, 2003). Applied to AI, this would imply that a boost from primitive to moderate abilities would be perceived as just as satisfying to users as a subsequent boost from high to exceptional, even though the latter is a bigger technical leap—this is in line with the observation that human ratings of language models grow sub-linearly with regard to model size (Li et al., 2022). Clark et al. (2021) also observed user satisfaction with AI assistants levels off prior to technical milestones being achieved, implying a psychological ceiling above which subjective value is reduced. This logarithmic trend creates a "satisfaction gap," where objective accomplishment outstrips user appreciation, an effect perhaps enhanced by the rapid pace of AI advancement. Polyportis (2024), for example, observed a drop in usage of ChatGPT across eight months as novelty wore off ($t(221) = 4.65$, $p < 0.001$), suggesting users have integrated its abilities into everyday expectations, thus reducing perceived value as time elapsed.

Population dynamics make this model even more complex, as adoption and persistence mirror varied user responses to AI capabilities. Drawing on Bass' (1969) diffusion model, adoption is driven by forces external to the system, such as marketing and social pressure from early adopters, a process well established in technology diffusion (Rogers, 2003). Persistence, however, is founded on satisfaction levels, with attrition increasing as expectations exceed performance (Bhattacherjee, 2001). User heterogeneity—early adopters, mainstream users, and latecomers—drives adaptation rates, with leaders potentially normalizing capabilities at a faster rate due to greater initial expectations (Rogers, 2003). Kim and Malhotra (2005) point out how experience anchors evaluation, and that early adopters, accustomed to rapid progress, may learn more rapidly than late adopters faced with entrenched systems. This heterogeneity means that satisfaction trajectories vary by segment, with late adopters potentially maintaining higher levels for longer due to lower initial expectations, a pattern indicated by research on AI tool adoption (Polyportis, 2024).

The model also investigates the impact of capability delivery—punctuated or continuous—on adaptation mechanisms. Fenn and Raskino (2008) contend that technologies are prone to hype cycles of rapid progress resulting in ephemeral enthusiasm, followed by disillusion, which is comparable to psychological responses to novelty (Hirschman, 1980). For artificial intelligence, punctuated progress, like dramatic model releases, might temporarily restore user satisfaction by surpassing current reference standards; however, this is frequently followed by rapid adaptation as users reset expectations (Sheldon et al., 2013). Continuous development, on the other hand, results in a more gradual but less spectacular decline in satisfaction, since incremental innovation does not disturb current expectations (Bhattacherjee & Premkumar, 2004). This result suggests that a successful pacing strategy could balance partial adaptation with occasional revival, a postulate that is intuitively built into successful technology deployment cycles (Rogers, 2003).

Intervention strategies become a key element of this approach, trying to offset adaptation's dampening influence. Psychological theory suggests that introducing variety or novelty can retard adaptation by resetting reference points (Sheldon et al., 2013), a strategy applicable to AI via new features or interaction styles. Norton et al. (2012) discovered that personalization—customization of experience to individual tastes—increases engagement, implying that adapting AI capabilities to suit could maintain satisfaction. Expectation management, e.g., under-promising to over-deliver to expected performance, fits with Expectation-Confirmation Theory (Oliver, 1980; Bhattacherjee, 2001), and could retard normalization. Social benchmarking—comparing one's experience to others—may also influence satisfaction, though its effectiveness may decline over time (Wilson & Gilbert, 2008). These strategies, promising as they are, are themselves vulnerable to adaptation, as repeated exposure blunts their effect (Frederick & Loewenstein, 1999), requiring a dynamic, adaptive approach.

This theoretical framework combines findings from computational economics (Tesfatsion, 2006) and cognitive modeling (Marsella & Gratch, 2009), applying these principles to deployment in artificial intelligence applications. This theory assumes that satisfaction follows the trajectory of a logarithmic curve in terms of capability, demonstrating the action of psychological adjustment mechanisms (Dehaene, 2003), and develops through typical phases: steep initial gains when capabilities surpass base levels, diminishing returns as comparisons increase, balance when progress has become routine, and spasmodic shocks from qualitative leaps (Fenn & Raskino, 2008). Speed of adaptation, diversity of users, and patterns of delivery all influence these phases, with the potential for intervention (Klapwijk & Van Lange, 2009). While heuristic in design but not empirical in support for this particular deployment, this model is underpinned by extensive literature claiming a plausible trajectory consistent with diminishing satisfaction with AI tools like ChatGPT (Polyportis, 2024) and conversational systems (Leviathan & Matias, 2018). This theoretical framework has a strategic function as a foundation to inform discussion, challenging stakeholders to AI to consider not only the scaling of capability, but also the mechanisms by which users accommodate progress.

**4. Illustrative Evidence from Theoretical Insights**

The dynamic interplay between increases in artificial intelligence (AI) capability and user satisfaction provides a rich soil for investigating the implications of hedonic adaptation, drawing conceptual leverage from both psychological and technology adoption theory. Hedonic adaptation, according to definitions by Brickman and Campbell (1971), is the belief that people rapidly adapt to major changes and rebound to a stable level of happiness—a phenomenon validated across a variety of fields including income increases (Easterlin, 1995), disability adaptations (Oswald & Powdthavee, 2008), and consumer satisfaction (Frederick & Loewenstein, 1999). In the domain of AI, this perspective assumes that users will rapidly adjust to even record technological advancement, generating a "satisfaction gap" between measurable advancement and personal experience. This section demonstrates how such adaptation might take place, speculating that satisfaction increases in a logarithmic function to increases in capability, transitions through separate phases, and is influenced by factors such as the adaptation rate, capability delivery patterns, diversity of users, and intervention approaches. These conclusions, based on current research, can provide theoretical leverage for developing the dynamics of AI adoption without necessitating immediate empirical justification.

One possible trend is that AI user satisfaction peaks first as new capability outperforms low expectations before leveling off as these gains become the new norm, following a logarithmic trend in line with psychophysical principles. Subjective perception is claimed to increase logarithmically with stimulus intensity by the Weber-Fechner law (Dehaene, 2003), a principle transferable to technology where initial progress—i.e., from rudimentary chatbots to GPT-3 text generation (Brown et al., 2020)—may charm users as much as subsequent progress to GPT-4 accuracy rises (Bodonhelyi et al., 2024), without the latter surpassing the former as much technically. Sheldon et al. (2013) saw similar diminishing return in technology-fueled happiness, a phenomenon where each additional capability increase brings smaller subjective gains. Empirical evidence comes from Bodonhelyi et al. (2024), who report that users favored GPT-3.5's original requests over GPT-4 reformulated requests despite improved metrics (t = 2.9821, p = 0.0067), perhaps reflecting normalization of expectations towards previous standards. Polyportis (2024) also saw a significant decline in ChatGPT usage after eight months (t(221) = 4.65, p < 0.001), as novelty initial phase lost steam, following a pattern of initial rapid gains followed by stabilization. This pattern follows Gartner's "hype cycle" (Fenn & Raskino, 2008), where technologies move from overhyped enthusiasm towards a plateau, a phenomenon that AI users may see early spikes, diminishing return, and irregular plateaus as capabilities become part of their baseline expectations. Figure 1 shows this theoretical trend with AI capability rising sharply on the x-axis while satisfaction followed a logarithmic pattern with increasing phases of initial rapid gains, diminishing return, plateaus, and temporary spiking, following observed adoption patterns.

The pace at which users adapt to AI developments likely drives this pattern of satisfaction, with psychological evidence demonstrating that adaptation velocity governs affect persistence. Frederick and Loewenstein (1999) argue that faster adaptation shortens return to baseline happiness, a mechanism observed in technology contexts where satisfaction reduces after uptake despite continued performance (Bhattacherjee & Premkumar, 2004). Kim and Malhotra (2005) theorize that earlier experiences anchor later judgments, indicating that users experiencing rapid

AI progressions—such as consecutive language model upgrades (Hoffmann et al., 2022)—will normalize capability faster, experiencing more catastrophic declines in satisfaction. Slower adapters, encountering mature systems later, may experience higher satisfaction longer before settling into a typical baseline, a mechanism observed in longitudinal technology studies (Bhattacherjee & Premkumar, 2004). Lyubomirsky (2011) argues that quick change pace accelerates reference point change, a mechanism feasible in AI with breakthroughs like multimodal models (Radford et al., 2021) outpacing user accommodation, hypothesizing short-term divergence but long-term equivalence among adaptation velocities. This aligns with evidence showing conversational AI users quickly berate constraints over basking in achievements (Leviathan & Matias, 2018, p. 14), which highlights adaptation in tempering subjective responses. As shown in Figure 2, satisfaction paths vary hypothetically with adaptation speed—slower adapters enjoy higher levels longer before converging with faster adapters, a mechanism observable in psychological processes of adjustment.

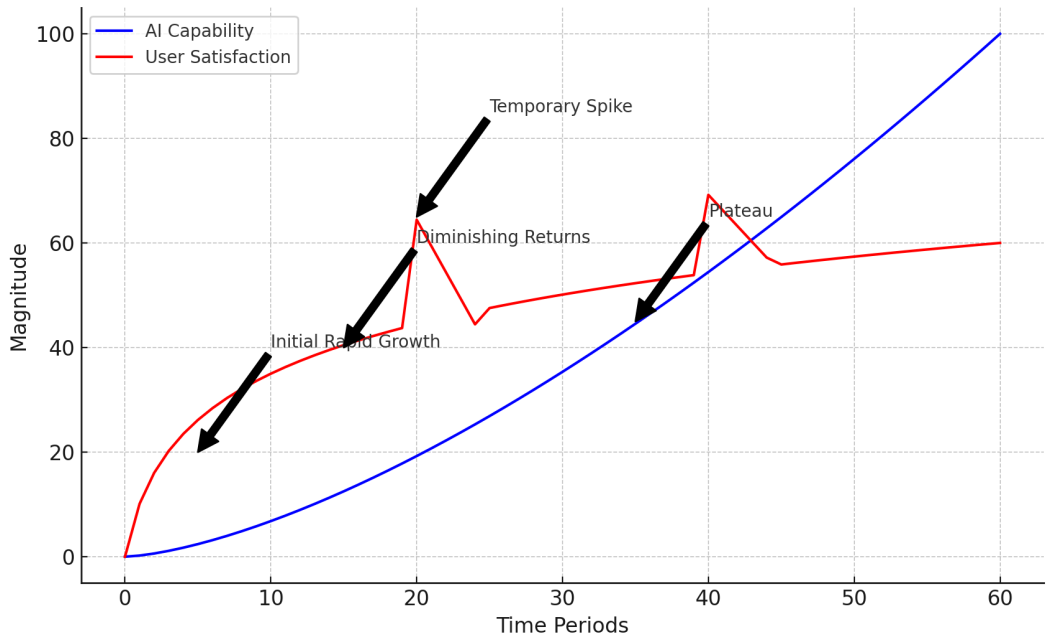

**Figure 1: Hypothetical Capability vs. Satisfaction Over Time**

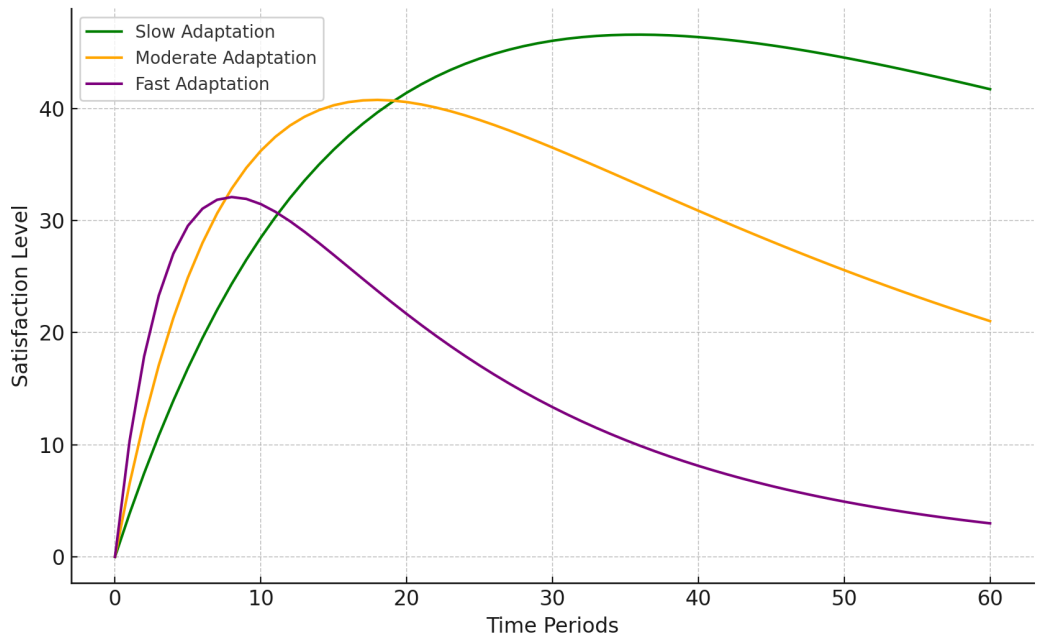

**Figure 2: Hypothetical Satisfaction Trajectories by Adaptation Pace**

Capability delivery patterns—either continuous or punctuated—can further elucidate these dynamics, with theoretical accounts suggesting differentiated satisfaction responses. Fenn and Raskino (2008) explain how sudden technological breakthroughs create fleeting excitement and subsequent disillusionment, a cyclical pattern consistent with psychological responses to novelty (Hirschman, 1980). For artificial intelligence, punctuated upgrades in the form of major releases introducing new capabilities (Wei et al., 2022) can temporarily boost satisfaction by outperforming current reference points, only to be followed by rapid adaptation as users acclimatize (Sheldon et al., 2013). For instance, the leap from GPT-3 to GPT-4 may temporarily re-ignite user interest; however, Bodonhelyi et al. (2024) discovered rapid normalization. Continuous upgrades, for instance, in the form of iterative fine-tuning (Kaplan et al., 2020), are likely to lead to a more gradual decline in satisfaction, as incremental upgrades do not violate mature expectations, a pattern in technology adoption where gradual improvements become normalized (Bhattacherjee & Premkumar, 2004). Rogers (2003) suggests that timely product releases can maintain consumer interest, and that an optimal lag time—potentially balancing partial adaptation with rejuvenation—may maximize subjective effect, a strategy inherent in successful consumer technology cycles. Figure 3 illustrates these hypothetical outcomes, showing that continuous upgrades lead to a gradual decline in satisfaction, while punctuated upgrades create fleeting spikes followed by steeper declines, consistent with the transient effects of novelty.

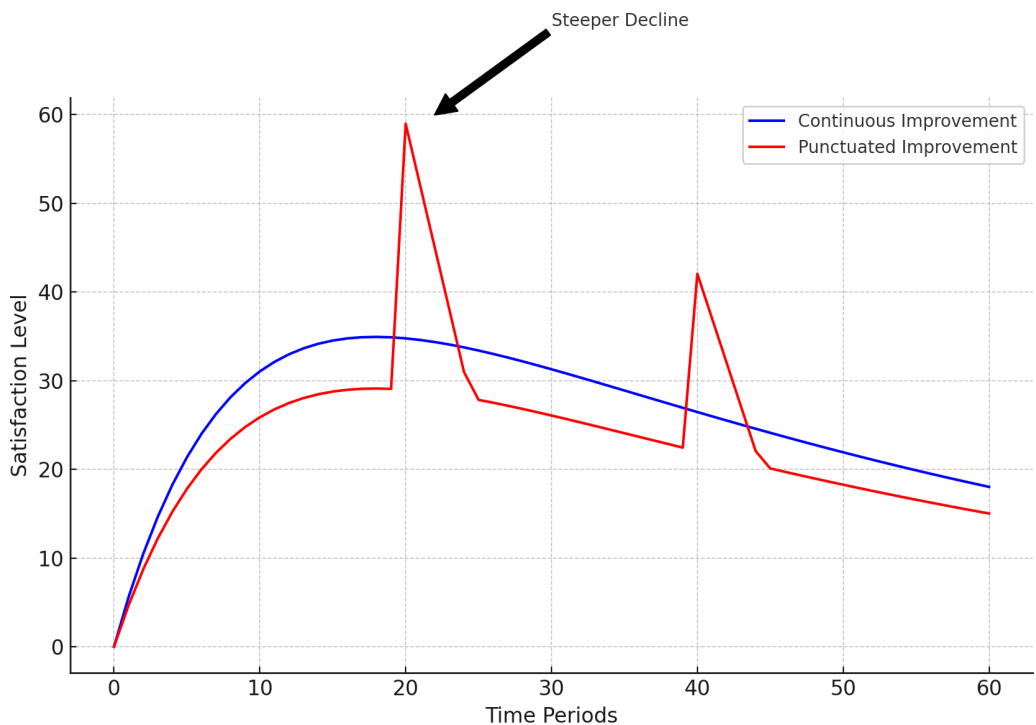

**Figure 3: Hypothetical Satisfaction Under Continuous vs. Punctuated Improvements**

User heterogeneity introduces a further variable, with adoption studies indicating that different segments have divergent satisfaction trajectories. Rogers' (2003) theory of diffusion describes early adopters, mainstream users, and late adopters, with different expectation profiles. Early adopters, often expectation-laden and technically sophisticated, can learn quickly, crest early but then decline as capabilities become routine—a pattern suggested by Polyportis' (2024) study of ChatGPT, where early users exhibited accelerated declines in usage. Mainstream users, with middle-of-the-pack expectations, could decline consistently, while late adopters, encountering mature systems with lower baselines, may persist longer on satisfaction (Rogers, 2003). Kim and Malhotra (2005) note that experience affects reference points, suggesting early adopters' accelerated exposure widens their satisfaction gap from newcomers, a dynamic seen in studies where sophisticated AI systems elicit more favorable responses from newcomers than veterans (Polyportis, 2024). This segmentation predicts no single capability rollout is suitable for all, with early adopters potentially more at risk of disengagement as expectations escalate (Bhattacherjee, 2001), a challenge encountered in AI application domains. Figure 4 illustrates these hypothetical trajectories, with early adopters cresting early and then declining, mainstream users declining consistently, and late adopters persisting, mirroring diffusion theory's adopter profiles.

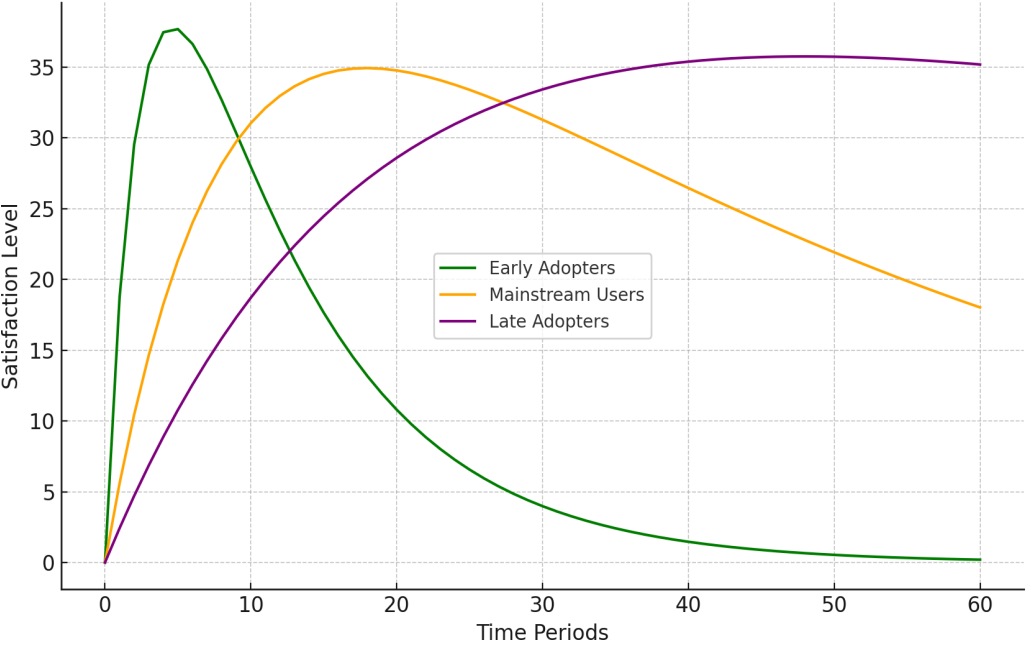

**Figure 4: Hypothetical Satisfaction Trajectories Across User Segments**

Intervention techniques offer theoretical means to slow the dampening impact of adaptation, drawing on consumer and psychological studies. Sheldon et al. (2013) propose variety or novelty can be used to retard adaptation by resetting reference points, a technique potentially applied to AI through new features or interaction modes—think going from text-based to multimodal output (Henighan et al., 2020). Norton et al. (2012) found personalization engages more, and it is feasible that tailoring AI experiences to personal taste can prolong satisfaction, as has occurred with consumer electronics successes like bespoke interfaces. Expectation management, such as under-promising then over-delivering on performance expectations, aligns with Expectation-Confirmation Theory (Oliver, 1980; Bhattacherjee, 2001), perhaps slowing normalization by maintaining low reference points. Social benchmarking—comparing against others—might also modulate satisfaction, but Wilson and Gilbert (2008) caution its impact may weaken with familiarity. These interventions are not, however, adaptation-proof; Frederick and Loewenstein (1999) note repeated use dulls novelty's impact, a weakness evident in technology where feature fatigue can result (Sheldon et al., 2013). Stacking interventions—like novelty and personalization—may exhibit synergies, a technique validated by consumer psychology (Norton et al., 2012), suggesting a multi-pronged approach to sustaining subjective value. As can be seen in Figure 5, such proposed interventions as feature novelty and personalization sustain satisfaction longer than expectation management or social benchmarking, suggesting multi-pronged approaches to resisting adaptation.

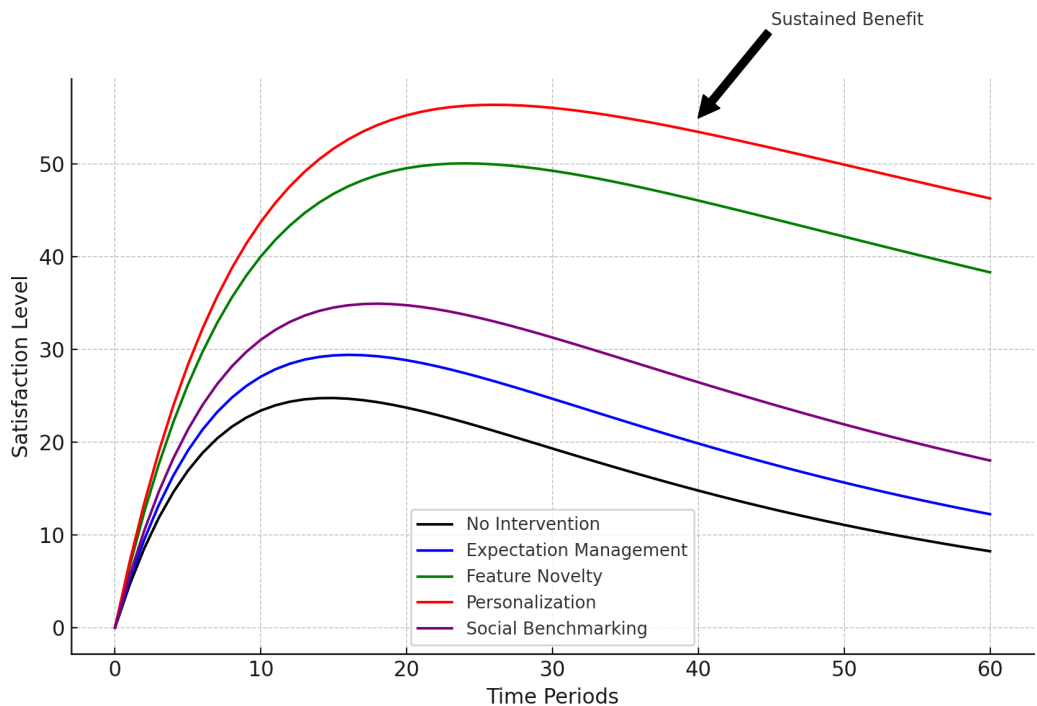

**Figure 5: Hypothetical Effects of Interventions on Satisfaction**

The theoretical models discussed above collectively indicate that satisfaction with artificial intelligence follows a logarithmic curve, moving through several stages: fast early gains as capabilities move beyond minimal levels, diminishing returns as standards increase, a phase of stabilization where improvements become normal, and the occasional bounds achieved through major breakthroughs (Fenn & Raskino, 2008). Rate of adaptation determines longevity, with quicker adjusters reaching ceilings earlier (Kim & Malhotra, 2005), and delivery patterns—punctuated spikes versus steady flow—determine peak intensity (Hirschman, 1980). User heterogeneity makes it hard, with early adopters falling off most rapidly and latecomers holding on longest (Rogers, 2003), and interventions relaxing temporarily (Sheldon et al., 2013). Supported by evidence in the form of waning ChatGPT enthusiasm (Polyportis, 2024) and language model preference shift (Bodonhelyi et al., 2024), this model explains how hedonic adaptation can buffer AI's subjective impact, requiring strategic re-examination above and beyond technical scaling.

## 5. Practical Implications

Exponential growth of artificial intelligence (AI) ignited by scaling laws between performance and computation, data, and model parameters (Kaplan et al., 2020) presents a paradox: technical capability grows exponentially but user satisfaction often lags behind. This can be understood through the lens of the hedonic adaptation process (Brickman & Campbell, 1971). This psychological process includes a response to revolutionary change, followed by reaching a state of happiness equilibrium (Frederick & Loewenstein, 1999), which implies incremental AI development will be unable to sustain perceived worth without deliberate intervention. In drawing on relevant research in psychology and technology adoption, this section outlines the pragmatic implications for AI development practice, user experience design, market forces, and ethics. Such implications require a shift from single technical optimization and towards embracing adaptation buffer effects approaches in seeking to trade off AI potential against long-term human satisfaction across environments. For AI roadmaps to development, stepwise incremental gain inducing diminishing subjective returns curve of ability and satisfaction (Dehaene, 2003)—logarithmic curve of ability and satisfaction—corresponds to the incremental advance paradigm along familiarity axes. Rather than continuously optimizing current measures, e.g., language model performance (Hoffmann et al., 2022), developers can seek to set out to achieve qualitative jumps in priority that knock user expectations over, e.g., emergent capability in large models (Wei et al., 2022). Psychological theory anticipates novelty resetting reference points, slowing adaptation (Sheldon et al., 2013), so introducing radically new capability—e.g., from text to multimodal output (Radford et al., 2021)—can rekindle user enthusiasm more than incremental gain. Fenn and Raskino (2008) note how punctuated technology development creates temporary enthusiasm, so careful timing of release, allowing partial adaptation between jumps, can maintain satisfaction more than gradual upgrade or sporadic overhaul. Rogers' (2003) diffusion theory also anticipates early emphasis on slower-adapting user segments—those less subjectively exposed to rapid normalization (Kim & Malhotra, 2005)—to establish a stable platform for satisfaction, while serial release of advanced capability to new users with lower baselines maximizes impact, since new adopters rate mature systems higher (Polyportis, 2024). Technical benchmarks, too, must account for more than raw measure of performance, with emphasis on new capability rather than routine improvement, since Diaz and Madaio (2023) argue that current tests misalign with heterogenous user perception, perhaps overemphasizing incremental achievement users soon take for granted (Lyubomirsky, 2011).

User experience design offers a different path to counter adaptation, applying psychological principles to manage reference points and prolong satisfaction. Klapwijk and Van Lange (2009) propose that satisfaction is reference-dependent, and incremental revelation—revelation of capabilities in a sequence rather than all at once—would slow normalization by keeping expectations in disequilibrium, a strategy shown effective in consumer technology (Sheldon et al., 2013). Introducing new ways of interacting, even with homogeneous underlying technology, can restart adaptation cycles, as variety keeps things interesting (Frederick & Loewenstein, 1999); e.g., shifting how users interact with an AI assistant, from text to voice commands (Leviathan & Matias, 2018), may revive interest. Personalization, tailoring capabilities to individual tastes, increases perceived value (Norton et al., 2012), so adaptive AI systems—tuning outputs to user habits—may prolong satisfaction longer than homogeneous installations, a strategy shown effective in personalized consumer products (Yang & Forney, 2013). Paradoxically, prospect theory's asymmetry (Kahneman & Tversky, 1979) predicts that intermittent performance drops followed by recovery may increase satisfaction by exploiting loss aversion—users feel greater relief at restored capability than pleasure at consistent gains—a strategy raised in user tolerance research on tech bugs (Bhattacherjee, 2001). These psychological principle-driven design strategies move away from maximizing capability exposure towards strategically pacing and shaping user interaction to counter adaptation's irresistible force.

The adaptive power reconfigures competitive strategies and market forces, as diminishing returns on established capabilities create new opportunities and challenges. As Rogers (2003) states, early adopters demonstrate rapid adaptation, while pioneering firms—those that bring innovations such as GPT-3 (Brown et al., 2020)—are susceptible to rapid normalization of user expectations, which can reduce satisfaction levels regardless of their technological lead. Followers, however, can take advantage of the lower reference points of new users, obtaining the same satisfaction level through lower-technology systems; this is what one finds in the success stories of late-entrant technologies (Schmidt & Cohen, 2017). This competitive pressure triggers innovation, as diminishing returns on updates to existing features—seen in incremental updates to language models—force firms to seek qualitatively new capabilities in order to trigger new cycles of adaptation (Brynjolfsson et al., 2018). This phenomenon explains the cycles in mature artificial intelligence markets, where new applications dominate over incremental improvement. Market segmentation emerges, as different capability portfolios are appropriate for different adaptation profiles—cutting-edge novelty can be pursued by early adopters, while later adopters may pursue reliable utility (Polyportis, 2024)—which is consistent with the segmented strategy advocated by diffusion theory (Rogers, 2003). Also relevant is the frequency of use; Orben and Przybylski (2019) contend that sporadic use maintains perceived value better than constant use, and thus business models that allow periodic rather than continuous use—frequent subscription breaks or rotating features—can support adaptation, in contrast to always-on AI systems, which can lead to user fatigue (Lyubomirsky, 2011).

Ethics and policy issues arise when adaptation effects breach technical and commercial domains and challenge social impact issues for AI. If intangible benefits level out as use increases with increasing model size—such as computational burden of large models (Hoffmann et al., 2022)—this contradicts uncontrolled growth as possible, an issue resonating in enterprise AI critique (Davenport & Ronanki, 2018). Capability imbalance and familiarity can require disclosure, and Schmidt and Cohen (2017) suggest adaptation effect disclosures to facilitate consumer choice, such as warning labels in other tech fields. Sensitization regarding adaptation mechanisms is, however, vulnerable to exploitation—designers can use psychological bias for short-term advantage, such as addictive tech habits (Norton et al., 2012)—necessitating ethical checks or regulation to prevent such habits, an issue still being debated in digital ethics (Venkatesh et al., 2012). Iterative exposure to successive AI, and rapid adaptation, can also instill technostress or dissatisfaction spirals in which users fall behind despite progress (Lyubomirsky, 2011), and digital well-being issues in design become pertinent, such as intrinsic pauses or expectation moderators (Orben & Przybylski, 2019). These ethical issues result in AI design that views adaptation not merely as a technical issue but as a public issue, balancing innovation and human-focused objectives.

In combination, these implications necessitate a change of course for AI, from a sole emphasis on scaling resources to a more profound integration of psychological forces. Development approaches may transition to punctuated, qualitative leaps over continuous fine-tuning, riding novelty to maintain user engagement (Sheldon et al., 2013), while dominance of varied adoption phases maximizes happiness across segments (Rogers, 2003). User experience design may utilize progressive disclosure, personalization, and strategic variation to retard adaptation (Klapwijk & Van Lange, 2009), resisting logarithmic dampening of subjective gains (Dehaene, 2003). Competition in markets relies on dominance of adaptation's forces, benefiting innovators who reset expectations and segmenters who customize offerings (Brynjolfsson et al., 2018), with usage models avoiding overexposure (Orben & Przybylski, 2019). Ethically, transparency and well-being shields guarantee that scaling benefits users, not systems (Schmidt & Cohen, 2017). These approaches, drawing on psychology literature (Frederick & Loewenstein, 1999), adoption (Rogers, 2003), and AI critique (Diaz & Madaio, 2023), demonstrate how hedonic adaptation recontextualizes AI's practical landscape. They provoke stakeholders to transcend technical ability, aligning development with the human capability to appreciate it, ensuring the promise of AI is realized as lasting value and not transient wonder.

## 6. Limitations and Future Directions

The hedonic adaptation theory of AI adoption, as built here, provides a valuable contribution to explain the gap between technical competence and user satisfaction. The discussion is based on psychological hypotheses built by Brickman and Campbell (1971) and technology adoption theories presented by Rogers (2003). However, the use of theoretical synthesis by the framework rather than ambiguous, context-dependent data reveals some limitations that might constrain its application and provide fertile ground for future research. Hedonic adaptation, as proven in numerous domains, including income changes (Easterlin, 1995) and consumer experience (Frederick & Loewenstein, 1999), suggests that users quickly adapt to improvements in AI; however, the complexity of this effect across various AI domains remains to be explored. This section discusses these limitations—ranging from the lack of longitudinal specificity to overly simplistic capability models—and outlines research directions towards the calibration and expansion of this model as it seeks to account for the complexity of human-AI relationships in an era of fast-moving technology change.

A main limitation is the model's wide calibration with adoption and psychological literature, as opposed to specific longitudinal data on artificial intelligence that may measure adaptation trajectories accurately over time. Polyportis (2024) reports this phenomenon in an eight-month trial of ChatGPT usage, showing a dramatic decrease as early novelty wore off ($t(221) = 4.65$, $p < 0.001$); however, such experiments are rare and limited in scope to corroborate adaptation rates across different AI systems, such as language models (Brown et al., 2020) or vision tools (Zhai et al., 2022). Bhattacherjee and Premkumar (2004) reported declines in satisfaction after adoption in general technological environments; however, in the lack of longitudinal datasets dedicated to AI, the precise timing and duration of adaptation phases—initial benefits, diminishing returns, stabilization—remain speculative (Fenn & Raskino, 2008). Future research should aim at long-term monitoring of user satisfaction for specific AI tools, such as conversational agents (Leviathan & Matias, 2018) or decision-support systems, in order to disambiguate how adaptation occurs over months or years, thus providing empirical evidence to refine this model's assumptions on the logarithmic curve of satisfaction (Dehaene, 2003).

The model's generalizability also breaks down in most AI application contexts, where adaptation patterns and rates can differ significantly. Creative applications such as generative AI (Henighan et al., 2020) can be employed longer by users due to the variety of their output, decelerating adaptation as novelty endures (Sheldon et al., 2013), whereas decision-support systems, with regard for consistent utility, might normalize more quickly as users value reliability over innovation (Davenport & Ronanki, 2018). Psychological theory predicts adaptation is domain-specific—emotional reaction to hedonic vs. utilitarian rewards varies (Lyubomirsky, 2011)—but this model presumes a single process. Comparative studies across domains, from art AI to industrial optimization, can identify domain-specific adaptation processes, assessing whether creative applications are more immune to normalization than functional applications, as indicated by user persistence with diverse tech experiences (Norton et al., 2012). Such research would enhance the model, adapting its implications to specific AI environments.

User heterogeneity, though recognized, is only treated at a general level, constraining the model's accuracy in accounting for psychological and demographic factors in adaptation. Rogers' (2003) theory of diffusion differentiates between early adopters, mainstream users, and latecomers, proposing varying expectation baselines, but age, experience, or personality factors—identified as influencing technology responses (Venkatesh et al., 2012)—are under-explored here. Parasuraman (2000) proposes that "technology optimism" among early adopters may retard adaptation to new systems, while domain experts may assess AI in terms of professional rather than personal reference points, shifting satisfaction patterns (Srite & Karahanna, 2006). Polyportis (2024) alludes to such heterogeneity, with student users demonstrating differential ChatGPT usage, but more detailed demographic information—covering age, education, or cultural tech attitudes—is required to refine adaptation rate assumptions. Future research could use surveys or experiments to examine how such factors modulate adaptation, optimizing interventions such as personalization proposed by Norton et al. (2012)—to effectively target specific user profiles.

A further constraint is the model's scalar representation of AI ability, reducing the richness of systems in the world to a one-dimensional representation. Scaling laws map performance against resources (Kaplan et al., 2020), yet AI abilities such as language generation (Hoffmann et al., 2022), image classification (Zhai et al., 2022), or multimodal fusion (Radford et al., 2021) cover multifaceted feature spaces. Users can accommodate differentially to accuracy than to originality, Bodonhelyi et al. (2024) illustrate through shifts in preferences between GPT-3.5 and GPT-4 outputs (t = 2.9821, p = 0.0067). Frederick and Loewenstein (1999) contend adaptation is stimulus-type dependent, and a vector representation of capability could better capture the way users compromise between different AI attributes—speed, dependability, novelty—over time. Taking this to higher-dimensional spaces, perhaps via comparative study of multifaceted AI tools (Henighan et al., 2020), would support richer analysis, examining whether adaptation along one dimension (e.g., accuracy) outperforms another (e.g., interactivity), as user feedback shifts suggest (Clark et al., 2021).

Social forces, while treated through diffusion (Bass, 1969), are underdeveloped, overlooking the role of collective processes in shaping adaptation. Wilson and Gilbert (2008) suggest that social comparison—comparison of one's experience with others'—regulates emotional reactions, a mechanism that may increase or decrease AI satisfaction as users come together online or in groups (Leviathan & Matias, 2018). Kim and Malhotra (2005) report that social anchors influence technology judgments, meaning that peer feedback about AI tools such as ChatGPT may speed normalization if deficits predominate debate (Polyportis, 2024). More advanced modeling of social processes—perhaps incorporating network analysis or expectation-sharing research—may uncover how collective reference points emerge, testing whether social reinforcement slows or speeds adaptation, a direction in need of study with AI's social embedding (Schmidt & Cohen, 2017).

Cultural context is another restriction, since adaptation varies across cultures with different expectations around technology. Srite and Karahanna (2006) cite how tech acceptance is influenced by cultural values, implying baseline expectations—and adaptation rates—vary across the globe; tech-rich area users can normalize AI quicker than emerging markets users (Parasuraman, 2000). Lyubomirsky (2011) posits emotional baselines are culturally relative, so cross-cultural research would find substantial differences—maybe Western users normalize conversational AI quicker (Leviathan & Matias, 2018) than Eastern users appreciating different tech characteristics. Future work will have to cross cultural contexts, comparing AI adoption adaptation across regions to tune this model's parameters, as intimated by global tech adoption differentials (Venkatesh et al., 2012).

These boundaries map out a solid research agenda. Long-term longitudinal studies of single AI systems—e.g., generative models (Brown et al., 2020) or assistants (Clark et al., 2021)—would confirm adaptation phases, using Polyportis (2024). Reference point manipulations in experiments, with varying capability exposure or social cues, would measure mitigation strategies (Sheldon et al., 2013), while comparative analysis in a particular domain investigates contextual variance (Davenport & Ronanki, 2018). Creating adaptation-resistant designs—novelty-driven or custom (Norton et al., 2012)—and merging this model with more general acceptance models (Venkatesh et al., 2003) would close theoretical holes. As AI increases in everyday life, these limitations will be overcome to better understand human adaptation, such that development keeps up with subjective experience, not technical ability (Kaplan et al., 2020), an undertaking at the heart of AI's social promise.

## 7. Conclusion

The development of artificial intelligence (AI), driven by scaling laws foreseeing exponential improvements in performance through increased computational capability, data, and model complexity (Kaplan et al., 2020), marks a staggering feat of technological progress. Examples like GPT-3 (Brown et al., 2020), computer vision systems (Zhai et al., 2022), and multimodal models (Radford et al., 2021) demonstrate this rapid development, delivering capabilities once considered fanciful. But this perspective identifies a profound disconnect: the objective gains do not easily map to sustained subjective satisfaction, which gap is explained by the paradigm of hedonic adaptation (Brickman & Campbell, 1971). Psychological research in the field suggests that people quickly adapt to large changes—whether in terms of increases in earnings (Easterlin, 1995) or acquisition of new goods (Frederick & Loewenstein, 1999)—thus returning to a happiness baseline, a phenomenon that implies the constant breakthroughs of AI could be tempered by a logarithmic satisfaction curve where each step yields diminishing returns (Dehaene, 2003). This inference combines with the idea that adaptation revalues AI's value proposition to call for a strategic realignment to balance technical capability with sustained human experience.

The logarithmic curve, whereby subjective gains plateau despite exponential capacity gains, has profound implications for artificial intelligence's future, thereby denying the assumption that scaling leads to user satisfaction. Technology adoption studies show that post-adoption satisfaction decreases as initial euphoria wears off (Bhattacheerjee & Premkumar, 2004), a pattern seen in artificial intelligence technologies like ChatGPT, where usage slowed as novelty wore off (Polyportis, 2024). Leviathan and Matias (2018) demonstrated that users of conversational AI quickly shifted attention from marvels to limitations (p. 14), in line with psychological studies that show rapid change accelerates reference point shifts (Lyubomirsky, 2011). Such a process—seen in Gartner's hype cycle from euphoria to stabilization (Fenn & Raskino, 2008)—means that AI developers, designers, and policymakers need to broaden their vision beyond resource allocation to grapple with the mediating power of adaptation, an imperative adopted by research that shows human appraisal lags behind technical standards (Clark et al., 2021; Li et al., 2022). The stakes are high: as artificial intelligence increasingly penetrates daily life, unsolved satisfaction plateaus risk diminishing its social impact despite high technology investment (Brynjolfsson et al., 2018). The future trajectory does not call for abandonment of scaling but rather the embracing of adaptation-aware strategies that positively address the psychological undertones of human-AI interaction. Sheldon et al. (2013) indicate that novelty and variety factors can impede adaptation, referencing developmental strategies that concentrate on qualitative progress—such as emergent capabilities (Wei et al., 2022)—over incremental additions. User experience designs that regulate exposure to capabilities or tailor interaction (Norton et al., 2012) may maintain user engagement, while market strategies that address heterogeneous user segments (Rogers, 2003) and ethical frameworks that guarantee transparency (Schmidt & Cohen, 2017) balance innovation with user bliss. This view, based on theories of psychological adaptation (Frederick & Loewenstein, 1999) and insight into technology diffusion (Bass, 1969), represents a tentative move toward the alignment of human-centered factors within the technical debate around AI. As AI capabilities continue to increase, the disparity between its potential and the emotional experience it elicits will continue to widen unless proactively addressed, rendering adaptation-aware development an imperative model for the discipline (Kaplan et al., 2020). By recognizing that satisfaction is as much a human construct as capability is a technical construct, AI has the ability to transition from an impressive technological achievement to an agent of enduring value, thereby realizing its potential in a manner that speaks meaningfully to its users.

**Statements**

**Author Contributions:**

All the authors contributed equally at all the stages of research leading to the submission of the manuscript.

**Conflict of Interest:**

The authors declare no conflicts of interest related to this research.

**Software & AI Usage Statement:**

Grammarly was used for language correction. ChatGPT generated code for visualizations, which was manually refined and validated by the authors. The final analyses and interpretations were conducted and verified by the authors.

(2022). Emergent abilities of large language models. *arXiv preprint arXiv:2206.07682*. Retrieved from https://arxiv.org/abs/2206.07682

Wilensky, U. (1999). *NetLogo*. Center for Connected Learning and Computer-Based Modeling, Northwestern University. Retrieved from http://ccl.northwestern.edu/netlogo/

Wilson, T. D., & Gilbert, D. T. (2008). Explaining away: A model of affective adaptation. *Perspectives on Psychological Science, 3*(5), 370–386.

Yang, S., & Forney, J. C. (2013). The moderating role of consumer technology anxiety in mobile shopping adoption: Differential effects of facilitating conditions and social influences. *Journal of Electronic Commerce Research, 14*(4), 334–347.

Zhai, X., Kolesnikov, A., Houlsby, N., & Beyer, L. (2022). Scaling vision transformers. *Proceedings of the IEEE/CVF Conference on Computer Vision and Pattern Recognition*, 12104–12113.
44